\documentclass[english]{emulateapj}
\usepackage{color}

\makeatletter

\@ifundefined{textcolor}{}
{%
 \definecolor{BLACK}{gray}{0}
 \definecolor{WHITE}{gray}{1}
 \definecolor{RED}{rgb}{1,0,0}
 \definecolor{GREEN}{rgb}{0,1,0}
 \definecolor{BLUE}{rgb}{0,0,1}
 \definecolor{CYAN}{cmyk}{1,0,0,0}
 \definecolor{MAGENTA}{cmyk}{0,1,0,0}
 \definecolor{YELLOW}{cmyk}{0,0,1,0}
 }

\makeatother

\usepackage[normalem]{ulem}  
\newcommand{\new}[1]{{\color[rgb]{0.8,0,0}#1}}

\newcommand{\old}[1]{{\color[rgb]{0.4,0.7,0.4}\sout{#1}}}
\newcommand{\blue}[1]{{\color[rgb]{0,0,1}#1}}

\usepackage{babel}

\usepackage{natbib}

\shorttitle{A Neutron Star Oscillation During a Superburst}
\shortauthors{Strohmayer \& Mahmoodifar}

\begin{document}

\title{Discovery of a Neutron Star Oscillation Mode During a Superburst}

\author{Tod Strohmayer$^1$ and Simin Mahmoodifar$^2$ \\ {\normalfont 
$^1$Astrophysics Science Division and Joint Space-Science Institute, NASA's 
Goddard Space Flight Center, Greenbelt, MD 20771, USA} \\ {\normalfont 
$^2$Department of Physics and Joint Space-Science Institute, University of 
Maryland College Park, MD 20742, USA}}


\begin{abstract}

Neutron stars are among the most compact objects in the universe and
provide a unique laboratory for the study of cold ultra-dense matter.
While asteroseismology can provide a powerful probe of the interiors
of stars, for example, helioseismology has provided unprecedented
insights about the interior of the sun, comparable capabilities for
neutron star seismology have not yet been achieved. Here we report the
discovery of a coherent X-ray modulation from the neutron star 4U
1636$-$536 during the February 22, 2001 thermonuclear superburst seen
with NASA's {\it Rossi X-ray Timing Explorer} (RXTE) that is very
likely produced by a global oscillation mode. The observed
frequency is $835.6440 \pm 0.0002$ Hz (1.43546 times the stellar spin
frequency of 582.14323 Hz) and the modulation is well described by a
sinusoid ($A + B\sin(\phi - \phi_0)$) with fractional
half-amplitude of $B/A = 0.19 \pm 0.04 \%$ ($4-15$ keV). The
observed frequency is consistent with the expected inertial frame
frequency of a rotationally-modified surface g-mode, an interfacial
mode in the ocean-crust interface or perhaps an r-mode.  Observing an
inertial frame frequency--as opposed to a co-rotating frame
frequency--appears consistent with the superburst's thermal emission
arising from the entire surface of the neutron star, and the mode may
become visible by perturbing the local surface temperature.  We
briefly discuss the implications of the mode detection for the neutron
star's projected velocity and mass.  Our results provide further
strong evidence that global oscillation modes can produce observable
modulations in the X-ray flux from neutron stars.



\end{abstract}

\section{Introduction}

Neutron stars provide natural laboratories for the study of
ultra-dense matter.  A primary method for such studies is to
accurately measure the mass-radius relationship for neutron stars,
which depends directly on the equation of state of dense
matter. However, different phases of ultra-dense matter can have
similar equations of state, such that mass-radius measurements alone
may not be definitive in probing the composition of matter at the
highest densities \citep{2013PhRvD..88h3013A,2012ARNPS..62..485L}. Additional, complementary observables sensitive to the phase of
dense matter would be extremely valuable for a more comprehensive
understanding of neutron star interiors.  The frequencies and damping
timescales of different global oscillation modes of neutron stars,
such as the Rossby waves (r-modes) and gravity modes (g-modes), which
have the Coriolis force and buoyancy as their respective restoring
forces, depend on their interior structure and composition \citep{2000ApJS..129..353Y,2012PhRvD..85b4007A}. Thus, observations
of the oscillation modes of a neutron star can, in principle, provide
such a complementary probe of its interior properties \citep{2011MNRAS.413...47P}.

The low mass X-ray binary (LMXB) 4U 1636$-$536 (hereafter, 4U1636) is
a well studied accreting neutron star binary with an
optically-determined orbital period of 3.8 hr, and is also a
well-known X-ray burst source \citep{1998A&A...332..561A,2002ApJ...568..279G}.  Oscillations at $\approx 582$ Hz are detected during some
of its thermonuclear bursts and are almost certainly spin modulation
pulsations \citep{1998ApJ...498L.135S}. This conclusion is further
supported by the detection of $582$ Hz pulsations during a portion of
the thermonuclear superburst observed from this source on 22 February,
2001 with the Proportional Counter Array (PCA) onboard RXTE
\citep{2002ApJ...577..337S}.  Superbursts are rare, energetic X-ray
flares observed from LMXBs that are likely caused by unstable
thermonuclear burning of a carbon-rich layer formed from the ashes of
normal Type I X-ray bursts.  The time evolution of the 582 Hz
pulsations during the superburst was entirely consistent with the
system's known orbital ephemeris, and modeling of the frequency drift
enabled some constraints on the neutron stars's projected orbital
velocity and its rest-frame spin frequency \citep{2002ApJ...577..337S,2006MNRAS.373.1235C}.

\begin{figure*}
\begin{center}
\includegraphics[scale=0.6]{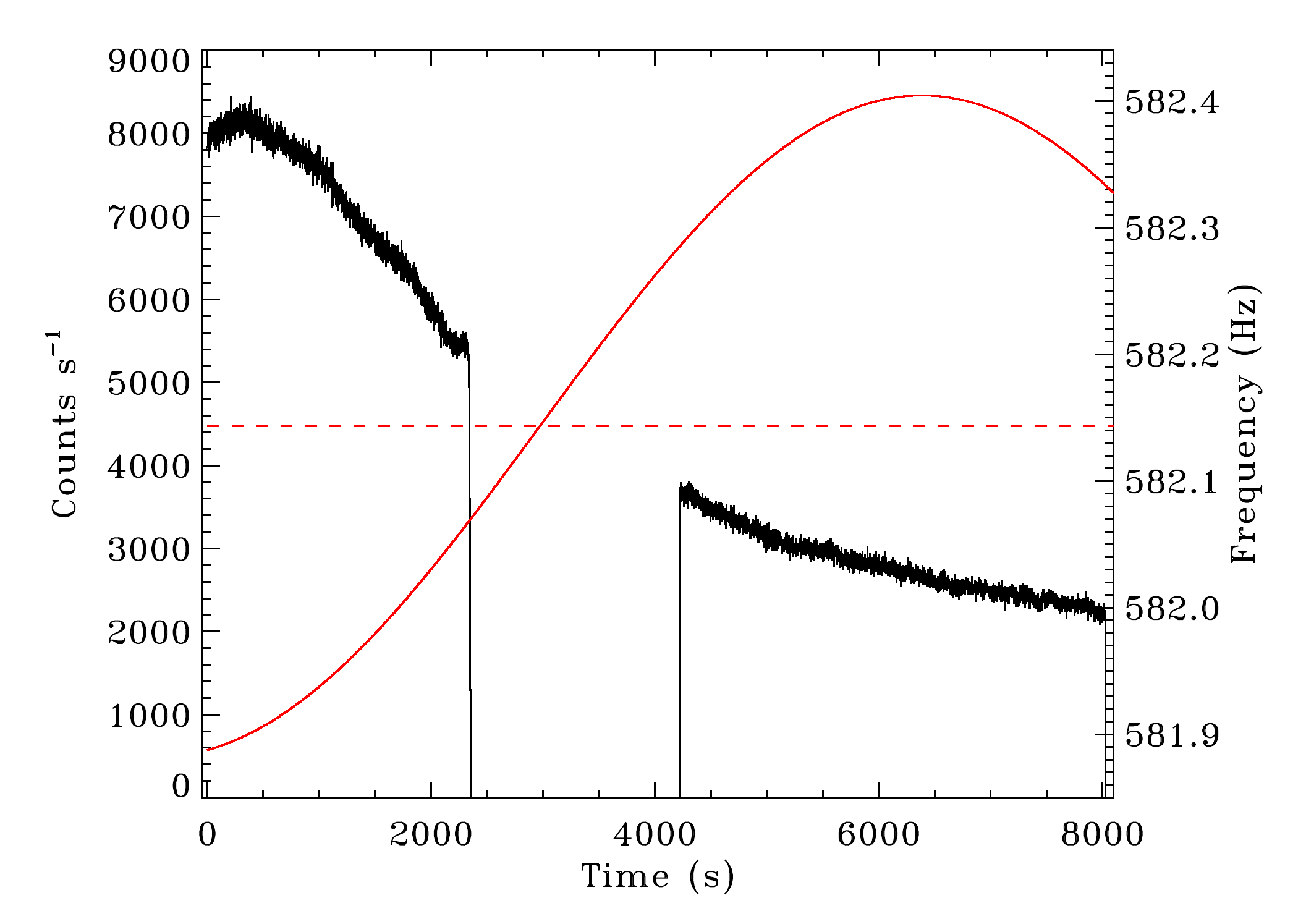}%
\end{center}
\caption{\label{fig:sb-lc} Light curve of the portion of the
superburst from 4U1636 used in our pulsation search, along with
the frequency evolution from our best-fitting orbit model (red
curve, right axis).  The data are the $4-15$ keV PCA count rates
(left axis) in 1 s bins.  The gap in the range from $\approx 2500-4000$ s results from Earth occultation of the RXTE satellite. The
first data interval begins just prior to the peak of the superburst.
Time zero is MJD 51962.70863263 (referred to the TDB timescale). }
\end{figure*}

The bright, hours-long superburst from 4U1636 provides a unique
opportunity to search for X-ray modulations produced by global neutron
star oscillation modes for several reasons. First, the combination of
its high luminosity (and thus high counting rate in the PCA) and long
duration provides excellent sensitivity to potentially weak
modulations. Second, the neutron star orbit constraint derived from
the 582 Hz pulsations enables a coherent search by removal of the
phase delays due to the neutron star's orbital motion \citep{2014ApJ...784...72S}. Finally, the thermonuclear burning provides a
plausible mechanism for the excitation of modes (the so-called
$\epsilon$-mechanism), and the shock wave produced by detonation of
the carbon fuel which powers the superburst is also likely to excite
wave motions at or near the surface \citep{1987ApJ...318..278M,
1996ApJ...467..773S, 2004ApJ...603..252P, 2012ApJ...752..150K}.

\begin{figure*}
\begin{center}
\includegraphics[scale=0.5]{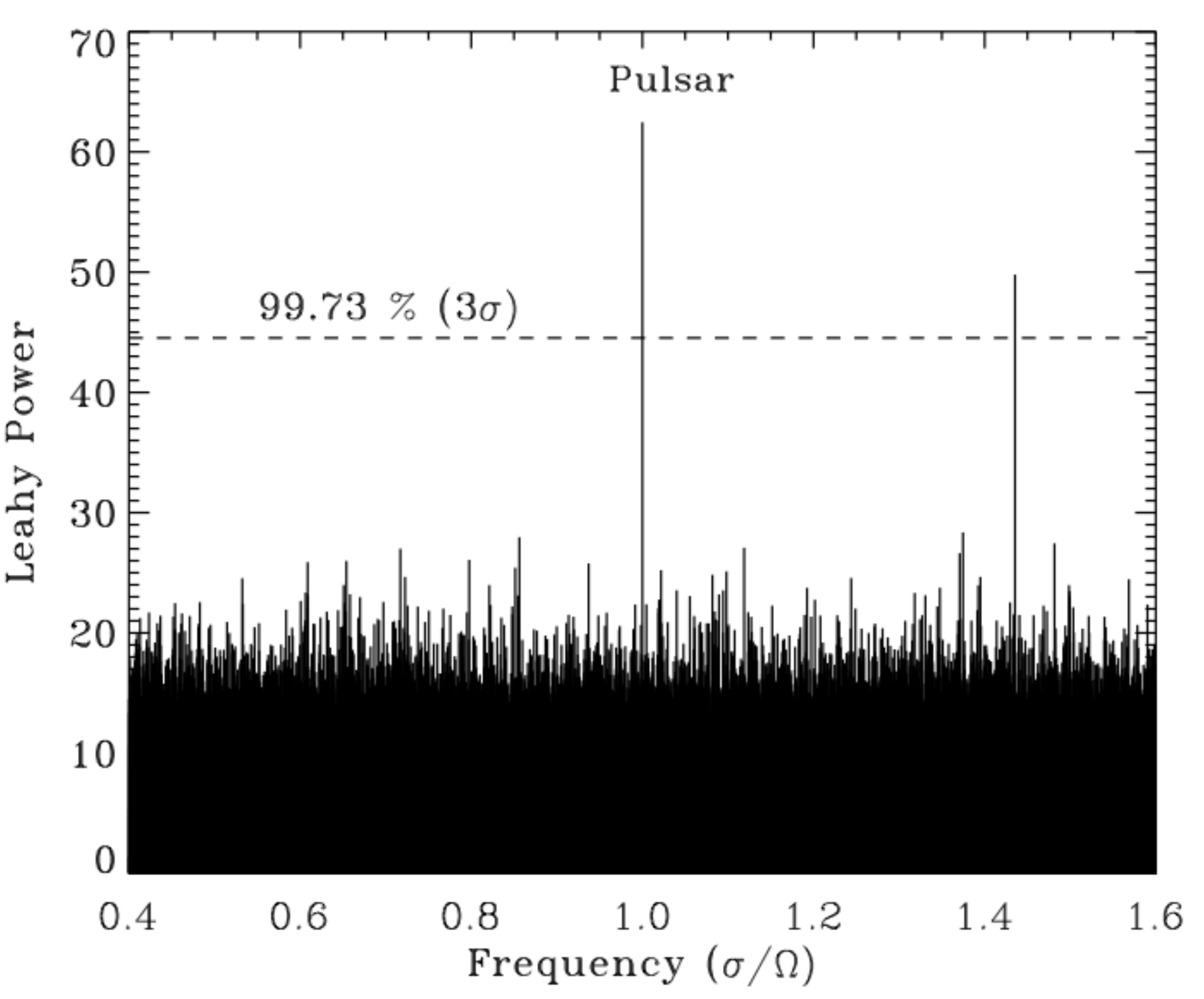}%
\end{center}
\caption{\label{fig:sbmode_pds} A portion of the power spectrum
(Leahy-normalized) computed from the 8000 s superburst interval shown
in Figure 1. The x-axis shows Fourier frequency scaled by the pulsar
spin frequency.  The plotted range includes both the co-rotating
($0.4166 \le \sigma/\Omega \le 0.7567$) and inertial ($1.243 \le
\sigma/\Omega \le 1.583$) frame frequency ranges searched.  The pulsar
signal appears at 1 in these units (labeled), and the oscillation mode
peak is evident near 1.43546, with a Leahy power close to 50.  The
dashed horizontal line denotes the $3\sigma$ detection level assuming
a number of trials equal to $4\times 3.17\times 10^6$. The factor of 4
conservatively accounts for the number of energy ranges searched.  }

\end{figure*}

\section{A Coherent Mode Search During the Superburst}

We carried out a coherent pulsation search during the superburst in
the manner described in detail in \citet{2014ApJ...784...72S}. As
the methods are clearly described there we omit some of the details
here. We used the high time resolution event mode
(E\_125us\_64M\_0\_1s) data obtained for most of the first two RXTE
orbits in which the superburst was observed (see Strohmayer \&
Markwardt 2002 for a discussion of the PCA data obtained during the
superburst).  For the orbit correction we began with the values that
minimized a $\chi^2$ fit to the phase timing residuals of the 582 Hz
pulsation, but with the orbit period fixed at the well-known value of
0.15804693(16) days (Strohmayer \& Markwardt 2002; Giles et
al. 2002). This gave a pulsar frequency of $582.14323$ Hz, a projected
neutron star velocity of 134.53 km/s, and an epoch of $T_{90} =
51962.74302977$ MJD (referred to Barycentric Dynamical Time, TDB).
These values are all consistent with those reported by Strohmayer \&
Markwardt (2002).  After correcting the event times for the neutron
star's orbital motion, we created light curves sampled at 4096 Hz
spanning 8000 s of the portions of the first two RXTE orbits with high
time resolution data (see Figure 1 in Strohmayer \& Markwardt
2002). We generated light curves in four energy bands, beginning with
the full PCA band, and then restricted the range to more closely match
that of the superburst's thermal emission component \citep{2014ApJ...789..121K}.  Our last and tightest energy range spans $\approx 4-15$
keV, where most of the detected superburst thermal emission
falls. This light curve re-sampled to 1 second bins is shown in Figure
1 along with the orbital frequency model used to correct the event
times (red curve).

We then computed power spectra and searched the same frequency ranges
as described in Strohmayer \& Mahmoodifar (2014).  These encompass the
expected range of frequencies--relative to the stellar spin
frequency--for r-modes and some rotationally modified g-modes, if a
mode modulates neutron star emission at either the mode's co-rotating
($0.4166 < \sigma/\Omega <0.7567$) or inertial frame frequency ($1.243
< \sigma/\Omega <1.583$). Here, $\sigma$ is a Fourier frequency and
$\Omega$ is the stellar spin frequency.  We note that for the
superburst's thermal emission component--which almost certainly
reflects emission from the whole surface of the neutron star--it would
seem the inertial frame frequencies are the more likely to be observed
\citep{2005MNRAS.361..504H,2005MNRAS.361..659L}.  This is in contrast to the case
of thermal emission from the hot-spot of an accreting millisecond
X-ray pulsar (AMXP), where the mode can periodically perturb the shape
of the hot-spot that is fixed in the rotating frame, producing an
observed modulation at the mode's co-rotating frame frequency \citep{2010MNRAS.409..481N}.

Our analysis revealed a significant power spectral peak at a frequency
of $835.644$ Hz ($\sigma/\Omega = 1.43546$), in the inertial frame
frequency search range.  The peak has a Leahy-normalized power of
49.3, and we estimate its significance as $\exp(-49.3/2) \times
N_{trials}$, where $N_{trials} = 4 \times 3.17\times 10^6$ is the
total number of Fourier frequency bins in the two ranges searched. The
factor of 4 conservatively accounts for the number of energy channel
ranges we used to compute light curves and power spectra, and results
in a significance of $2.5\times 10^{-4}$. The estimate is conservative
because the channel ranges searched were not all independent.  We also
carried out extensive Monte Carlo simulations to assess the
significance, taking into account the non-independence of the energy
channel ranges searched.  From these simulations we find a
significance of $\approx 1.5 \times 10^{-4}$, which gives us
additional confidence in the detection.  Figure 2 shows a portion of
the full power spectrum ($0.4 \le \sigma/\Omega \le 1.6$) which
includes the two frequency ranges searched, as well as the range
around the known pulsar frequency. There are clearly two significant
peaks present in the power spectrum, one is the known pulsar signal
(at 1 in these units), and the other is the putative oscillation mode
frequency at $1.43546$.  There are no other significant peaks
detected. As a further check, we used the Fourier frequency bins above
$\sigma / \Omega = 1.6$ to investigate the noise power distribution,
and find that it closely matches the expected $\exp(-P/2)$
distribution, giving us further confidence that our significance
estimates are robust.

The 835 Hz signal we detect is consistent with a coherent oscillation
over the 8000 s of the light curve. We emphasize that prior
searches, such as those which discovered the 582 Hz pulsations, would
not have detected this modulation because they did not account for the
motion of the neutron star, which smears a narrow band signal into
many adjacent Fourier frequency bins. An estimate of this effect can
be gleaned from Figure 1. The full (peak to trough) fractional change
in the frequency due to the orbital motion is $\Delta\nu/\nu \approx
0.5/582 = 8.6 \times 10^{-4}$. A coherent (single bin) signal at 835
Hz will be smeared into $N_f = 835 \times (\Delta\nu / \nu) / \delta
f$ bins, where $\delta f$ is just the frequency resolution of the
power spectrum, in this case $\delta f = 1 / 8000 \; {\rm s}$, and
$N_f = 835*(0.5/582)*8000 \approx 5700$. Any such signal would be
swamped by the poisson noise in this large number of Fourier bins
\citep{1991ApJ...379..295W}.

\begin{figure*}
\begin{center}
\includegraphics[scale=0.6]{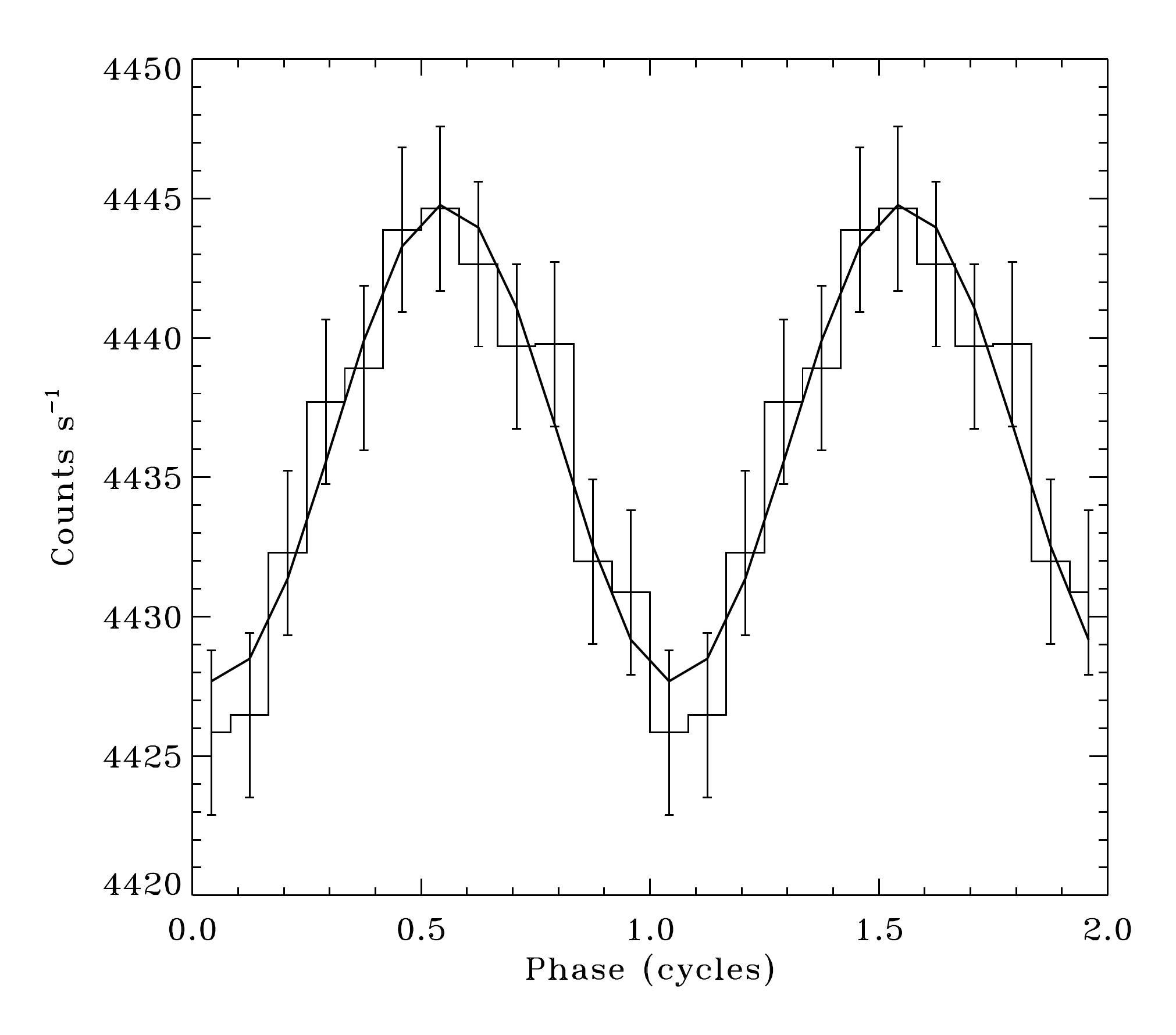}%
\end{center}
\caption{\label{sbmode_fold} Pulse profile in the $\approx 4-15$ keV
band obtained by folding the light curve data at the oscillation mode
frequency of $ \sigma/\Omega = 1.43546$ in 12 phase bins. Two cycles
are plotted for clarity. The best fitting sinusoidal model (solid
curve), $A + B\sin(\phi - \phi_0)$, is also plotted.  The
fractional half-amplitude of the best fitting model is $B/A =
0.19 \pm 0.04 \%$.  }
\end{figure*}

We folded the light curve data ($4-15$ keV) at the detected frequency
and fit a sinusoidal model of the form $A + B\sin (\phi - \phi_0)$ to
the resulting profile (see Figure 3). This model provides an excellent
fit to the data, and we find a fractional half-amplitude of $B/A
= 0.19 \pm 0.04 \%$. We do not find any evidence for significant
harmonics associated with this signal.  Assuming the observed
frequency is the inertial frame frequency of an oscillation mode
($\omega_i/ \Omega = 1.43546$), we can convert it to a co-rotating
frame frequency if we know its azimuthal wavenumber, $m$, since the
two frequencies are related as $\omega_i = m\Omega - \omega$.
Interestingly, for $m=2$ we find that $\omega / \Omega = 2 - 1.43546=
0.56454$, which is close to the candidate frequency ratio of 0.57276
we identified in the AMXP XTE J1751$-$305 (Strohmayer \& Mahmoodifar
2014), suggesting that the frequencies identified in the two sources
could perhaps be associated with the same oscillation mode.  For the
AMXP source the modulation mechanism is more likely due to
oscillation-induced perturbations to the hot-spot fixed in the
rotating frame, whereas for the superburst, emission from the entire
surface can be modulated by a mode, perhaps due to local variations in
the surface temperature. As mentioned above, these processes
would naturally lead to modulations at the co-rotating and inertial
frame frequencies, respectively.

\section{Discussion} 

Recent theoretical work has shown that the mode frequency ratio
relative to the spin frequency indicated for J1751$-$305 is
consistent, in principle, with the $l = m = 2$ r-mode for neutron
stars with plausible masses and radii \citep{2014MNRAS.442.1786A},
and surface g-modes with $l=1$ or 2 (Strohmayer \& Mahmoodifar
2014).  To the extent that an $m=2$ r-mode identification is
appropriate for the modulation during the superburst, and since the
mode frequency ratios are similar, the same r-mode identification
appears possible for 4U1636 as well. The $l=m=2$ r-mode is the most
unstable to the emission of gravitational waves and its amplitude can
grow exponentially--potentially causing a rapid spin-down of the
star--if viscosity and other non-linear dissipation processes are not
strong enough \citep{1998ApJ...502..708A,1998PhRvL..80.4843L, 1998PhRvD..58h4020O}.
The amplitude of the r-modes determines how fast they can
spin down the star. Now, the r-mode amplitude likely required to
account for the observed X-ray modulation during the superburst would,
if continuously present, produce a strong spin-down of the star
(Andersson et al. 2014; Strohmayer \& Mahmoodifar 2014).  Such a
spin-down is inconsistent with the pulse timing data for J1751$-$305
(Strohmayer \& Mahmoodifar 2014; Andersson et al. 2014), but we note
that there are no comparable long term spin-down measurements for
4U1636, as the source has never been detected as a persistent pulsar.
Excitation of the mode to an observable amplitude might also be
intermittent, perhaps driven by the superburst's energy release.

However, these inferences do not take account of the presence of a
solid crust within the neutron star. It has recently been argued that
the r-mode amplitude may be ``amplified'' in the surface layers above
the crust compared to the core because the crust can isolate the
surface layers from the core motions \citep{2014MNRAS.442.3037L}. Indeed, recent
estimates of r-mode amplitudes--at least those in the core--suggest a
low saturation amplitude of $10^{-7} - 10^{-8}$ \citep{2012MNRAS.424...93H, 2013ApJ...773..140M}.  Amplification of the mode
in the surface layers might then account for the observed X-ray
modulations, and if some process limits the r-mode saturation
amplitude in the core \citep{2013ApJ...778....9B}, then the
apparent inconsistency with the long-term spin evolution in
J1751$-$305 could perhaps be mitigated.  Alternatively, surface
g-mode or ocean-crust interface mode identifications remain viable,
and we note that rotational ``squeezing'' of the mode displacements
toward the rotational equator does not necessarily pose a problem for
the visibility of the mode in either of these cases, as the superburst
involves emission from the entire neutron star surface, and not from a
hot-spot localized near the rotational pole \citep{1996ApJ...460..827B, 2005ApJ...619.1054P}.

\begin{figure*}
\begin{center}
\includegraphics[scale=0.6]{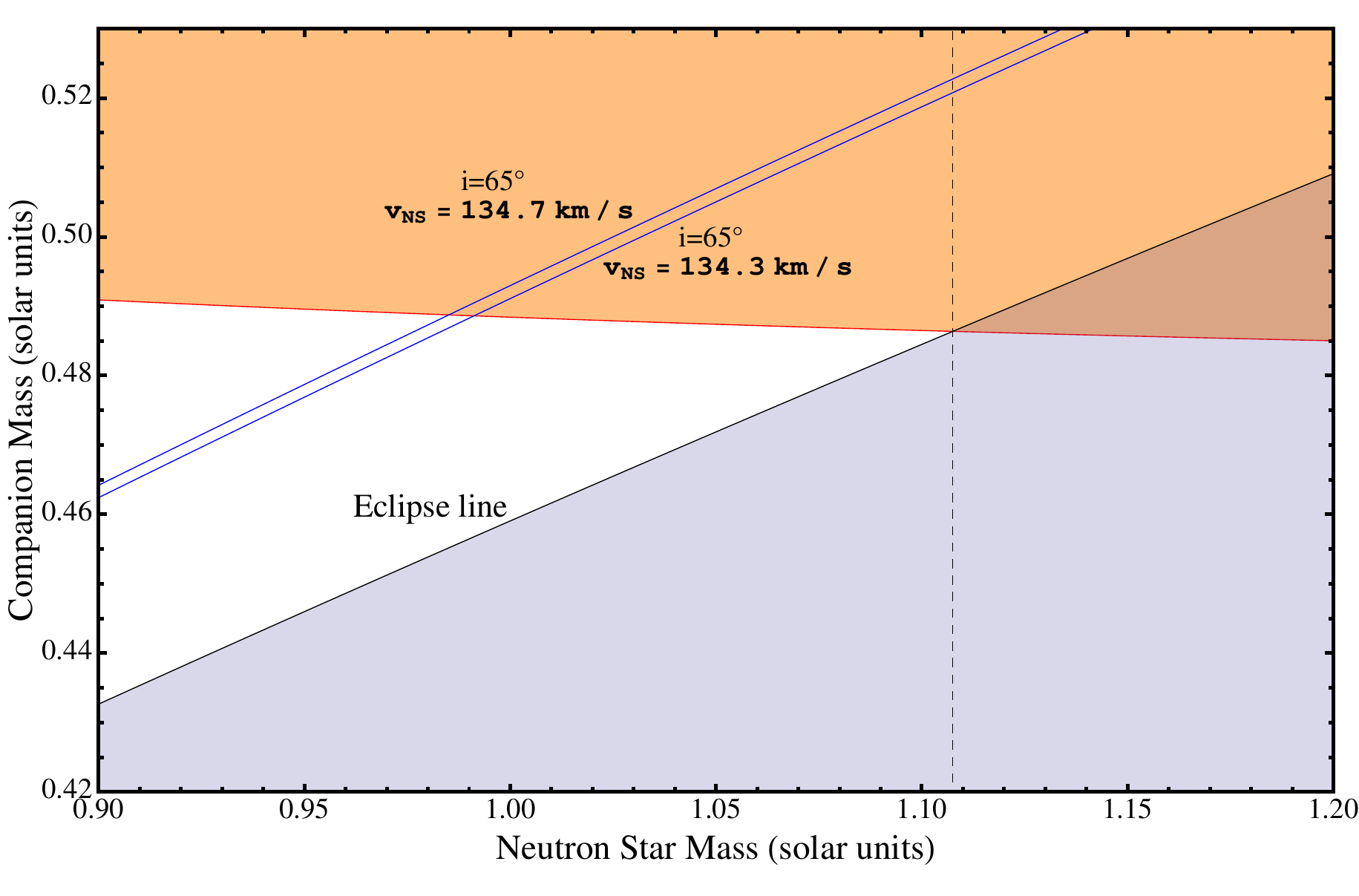}%
\end{center}
\caption{\label{fig:sbmode_massconstraints} Constraints on the
component masses of 4U1636 assuming the neutron star velocity of
$134.5 \pm 0.2$ km/s suggested by our mode frequency
detection. The orange region is excluded as zero age main sequence
(ZAMS) stars above this limit will have radii larger than the Roche
lobe radius for 4U1636. The blue region below the
``eclipse line'' is excluded as these systems would not be able to
satisfy the velocity constraint (taken as $v_{ns} \ge 134.3$ km/s) at an inclination angle which would preclude eclipses. The maximum allowed mass under
these assumptions is $M_{ns} \le 1.108 M_{\odot}$, and is marked by
the vertical dashed line. The region of allowed masses for a system inclination angle of $i =
65^{\circ}$ (inside the parallel blue lines) is also shown, to
indicate the precision which could be attained if both the inclination
and velocity were accurately known.}  
\end{figure*}

If the putative inertial frame frequency at $\omega_i/ \Omega =
1.43546$ is indeed constant over the 8000 s light curve interval, then
it suggests that the neutron star velocity can be much more
tightly constrained than indicated by the time evolution of the 582 Hz
pulsations alone (Strohmayer \& Markwardt 2002; Casares et al. 2006).
That is because the 582 Hz pulsations were detected over a much
  shorter fraction of the orbital period (800 s) than the 8000 s
  detection interval of the 835 Hz modulation. The much shorter pulse
  train of the 582 Hz oscillation was consistent with a much larger
  range of neutron star orbital velocity ($90-175$ km/s) than the 835
  Hz signal (Strohmayer \& Markwardt 2002). Indeed, varying the
orbital parameters away from the values for which the $835.644$ Hz
signal was detected results in a ``smearing-out'' of the signal
peak. As noted toward the end of \S 2, this is the expected
behavior of a coherent modulation when the chosen orbital parameters
do not accurately match the ``true'' values.

Assuming the mode frequency is constant in the star's frame, and using
it as a ``matched filter,'' we can, in principle, constrain the
projected neutron star velocity to $134.5 \pm 0.2$ km/s. A velocity
outside of this range results in significant signal loss in the
$835.644$ Hz oscillation. We caution, however, that this is a
somewhat indirect means to infer the star's velocity because we do not
have independent verification that the mode frequency is constant in
the star's frame. For example, some intrinsic variation of the mode
frequency could mean that the orbital velocity which maximizes the 835
Hz signal is different from the true orbital velocity.  Previous
radial velocity studies of 4U1636 based on both optical
spectroscopy and the 582 Hz superburst pulsations suggested a lower
projected neutron star velocity, in the range from $90-113$ km/s.
This was based on a ``spectroscopic ephemeris'' for the epoch (phase)
of superior conjunction of the companion deduced from Doppler
tomography of the Bowen emission lines observed from the optical
companion (V801 Ara; Casares et al. 2006).  The offset in the
``spectroscopic'' and ``photometric'' phases (used here in our
$835.643$ Hz signal detection) corresponds to $\approx 2.5\sigma$ of
the statistical uncertainty given for the reference epoch of the
spectroscopic ephemeris (Casares et al. 2006), however, the inclusion
of systematic uncertainties associated with identifying accurately the
site of Bowen fluorescence in the binary, as well as the (smaller)
uncertainty in the orbital period, would further reduce this potential
discrepancy.  We also note that the timing analyses of the 582 Hz
pulsations modestly favor---in yielding lower $\chi^2$ timing
residuals---the higher neutron star velocity suggested by our
$835.643$ Hz oscillation detection (Strohmayer \& Markwardt 2002).

Nevertheless, such a high neutron star velocity would have important
implications for its mass, as we now explain. 
Constraints on the component masses of 4U1636 assuming a neutron star
velocity of $134.5 \pm 0.2$ km/s are shown in Figure 4. Systems
within the orange shaded region are disfavored as any zero age main
sequence star (ZAMS) is larger than the Roche lobe radius for
4U1636. We used a standard estimate for the radius of the Roche lobe
\citep{1983ApJ...268..368E} and the stellar mass-radius relation of \citet{1996MNRAS.281..257T} to determine the lower boundary of this region.  The
``eclipse line'' denotes systems which give a neutron star velocity
equal to the $1\sigma$ lower bound of $134.3$ km/s at an inclination
angle for which eclipses just begin.  Since eclipses are not seen from
4U1636, the parameter space below this line is excluded. We used
Kepler's law and solved the Roche lobe eclipse geometry for a point
source \citep{1976ApJ...208..512C} to draw this line.
The allowed range of masses is also shown if the system inclination
were $i=65^{\circ}$ (within the parallel blue lines), and is meant to
suggest the accuracy that could be achieved if both the velocity and
inclination were well constrained.  The vertical dotted line marks the
maximum neutron star mass under these assumptions, $M_{ns} \leq 1.108
M_{\odot}$. From Figure 4 it is evident that an orbital velocity as
high as $134.5 \pm 0.2$ km/s would strongly favor a low mass
neutron star in 4U1636. Interestingly, there is growing
evidence for ``light'' neutron stars, including the X-ray binaries 4U
1538$-$52 and SMC X-1 \citep{2011ApJ...730...25R, 2013MNRAS.433..746C}. As
discussed by Lattimer (2012), the physics of core-collapse and the
thermodynamics of the resulting lepton-rich proto-neutron star,
suggests a minimum mass of about $0.9-1.1$ $M_{\odot}$ depending on
the entropy profile of the star \citep{1999A&A...350..497S}.
Thus, accurate measurements near the minimum would provide important
insights into the core-collapse physics \citep{2012ARNPS..62..485L}.  Finally, we
emphasize that an independent confirmation of the neutron star's
orbital velocity is crucial to definitively establish such a low mass
for 4U1636.

\section{Summary and Conclusions}

We carried out a search for high frequency coherent X-ray modulations
during the February 2001 superburst from the LMXB 4U 1636$-$536
observed with RXTE.  We used a circular orbit model deduced from the
time evolution of the 582 Hz pulsations seen previously during the
superburst to remove the Doppler delays due to binary motion of the
neutron star. We detected a coherent modulation at a frequency of
$835.6440 \pm 0.0002$ Hz (1.43546 times the stellar spin frequency of
582.14323 Hz), with a significance estimated from Monte Carlo
simulations of $\approx 1.5 \times 10^{-4}$, and a fractional
half-amplitude of $0.19 \pm 0.04 \%$. This frequency is consistent,
in principle, with the expected inertial frame frequency of an $l = m
= 2$ r-mode, a rotationally modified surface g-mode, or an ocean-crust
interfacial mode.  Our results provide further strong evidence that
global oscillation modes can produce observable modulations in the
X-ray flux from neutron stars. Sensitive searches for such signals
should be a priority for future, large area X-ray timing missions.

\acknowledgments We thank Cole Miller, Tony Piro and the anonymous
referee for helpful comments and discussions.
TS acknowledges NASA's support for high
energy astrophysics. SM acknowledges the support of the
U.S. Department of Energy through grant number DEFG02- 93ER-40762.

\bibliography{sb1636_ref}





\end{document}